\documentclass[onecolumn,amsmath,amssymb]{qm2008_abs}
\usepackage{epsfig}
\usepackage{graphicx}% Include figure files
\usepackage{dcolumn}% Align table columns on decimal point
\usepackage{bm}% bold math
\begin{document}

\title{{\Large Hot QCD equations of state and RHIC }}% Force line breaks with \\

\bigskip
\bigskip
\author{Vinod Chandra}
\email{vinodc@iitk.ac.in}
\author{Ravindra Kumar}
\author{V. Ravishankar}
\affiliation{Department of Physics,\\
Indian Institute of Technology Kanpur, Kanpur-208 016, India}
\bigskip
\bigskip

\begin{abstract}
\leftskip1.0cm
\rightskip1.0cm
We show how hot QCD equations of states can be adapted to make definite predictions for 
quark-gluon plasma at RHIC. We consider equations of state up to $O(g^5)$
 and $O[g^6(ln(1/g)+\delta)]$. Our method involves the
extraction of equilibrium distribution functions for gluons and quarks
from these equations of state by capturing the the interaction effects 
entirely in the effective chemical potentials. We further utilize these 
distribution functions to study the screening length in hot QCD and
dissociation phenomenon of heavy quarkonia states by combining this 
understanding with the semi-classical transport theory.

\vspace{1mm}
PACS: 25.75.-q; 24.85.+p; 05.20.Dd; 12.38.Mh\\
{\bf Keywords}: Hot QCD; equation of state; response function; quakonium dissociation; quark-gluon plasma
\end{abstract}
\maketitle
\section{Introduction}
The hot and dense matter created at RHIC behaves like a near perfect fluid as inferred from recent experimental observations
at RHIC\cite{star}. This form of the matter lies in the strongly interacting domain of QCD. It is of the interest to check if the domain is genuinely non-perturbative, or could be understood in  the form of  higher order terms in perturbation theory. Methods based on weak perturbative techniques \cite{weakc} have achieved partial success in interpreting the results from RHIC. The aim of the present article is to investigate more fully, the viability 
of pQCD EOS in understanding the RHIC results. We implement this programme by extracting quasi-free equilibrium distribution functions from the EOS
, capturing all the interaction effects in the  effective fugacities\cite{chandra1,chandra2}. This procedure
 allows one to study various observables for hot and dense matter at RHIC. 

\section{Quasi-particle realization of HOT QCD EOS}
We consider  hot QCD EOS in pQCD up to $O(g^5)$\cite{arnold}(EOS1) and $O[g^6(\ln(1/g)+\delta)]$\cite{kaj1}(EOS2). 
The free parameter $\delta$ in EOS2 encodes the unknown perturbative and non-perturbative contributions of $O(g^6)$ and can be fixed by matching the EOS with lattice predictions on EOS\cite{fkarsch}. 
The ansatz for the determination of equilibrium distribution functions involves retaining the ideal distribution forms, 
with the effective fugacities ($z_g=\exp(\mu_{g})$ and $z_f=\exp(\mu_f)$). 
Note that for the mass less quarks ( $u$ and $d$) which we consider to constitute the bulk of the plasma, $z_{g/f}=1$ ($\mu_{g/f}=0$) if they were not interacting. 

As the  first step in our approach, we express  the hot EOS in the  form
\begin{equation}
  P=P^I_g +P^I_q + \Delta P_g + \Delta P_{f}
\end{equation}
We seek to absorb the contributions from all the non-ideal terms in the effective chemical potentials $\mu_g$ and $\mu_f$. Since the equations of states are expected to be valid at $T>2T_c$,  we treat the dimensionless quantity $\tilde\mu_{g,f}  \equiv \beta \mu_{g,f}$ perturbatively. This determination of 
$\tilde\mu_{g/f}$ needs to  be done self consistently. Accordingly, we expand the  the grand canonical partition functions $(Z_{g/f})$ for gluons and quarks as a  Taylor series in $\tilde\mu_{g,f} $. We obtain
\begin{eqnarray}
\label{eqn7}
\log( Z_ g) = \sum_{k=0}^\infty (\tilde\mu_g)^k \partial^{k}_{\tilde\mu_g} \log(Z_g)|_(\tilde\mu_g=0) ;
\log(Z_q) = \sum_{k=0}^\infty (\tilde\mu_f)^k \partial^{k}_{\tilde\mu_f} \log(Z_f)|_{(\tilde\mu_f=0)}.
\end{eqnarray}
\\

We equate  $\log(Z)$ with the pressure by the well known relation :
$\log(Z)=P\beta V$,  and determine $\tilde\mu_{g/f}$ self-consistently order by order. We find that it is sufficient to determine them up to cubic order. Details may be found in \cite{chandra1}. 

\section{The screening length}
The Debye mass from equilibrium distribution function can be obtained from the expression \cite{kelly}
\begin{equation}
M^2_{D_{g/f}}= (g^\prime) 2C_{g/f}\int \frac{d^3p}{2\pi^3} \partial_{p^0} f^{g/f}_{eq} ,
\end{equation}
where $C_g=2N_c$, $C_f=-2N_f$ and $g^\prime$ is the effective coupling which appears in the transport theory. We fix $g^\prime$ by comparing the temperature dependence of the screening length with the lattice results of Kazmarek and Zantow\cite{zantow}. The behavior of the screening length for pure gauge theory and full QCD for EOS1 and EOS2 are shown in Fig.1. Interestingly, the behavior of the screening length with temperature qualitatively matches with the lattice QCD results(see Fig.2 of Ref.\cite{zantow}). The agreement for EOS2 is slightly better as compared to EOS1 and in both the cases agreement becomes better at higher temperatures($T\equiv 3T_c$ and higher).
\begin{figure}[htb]
\label{fig1}
\vspace*{-45mm}
\hspace*{-35mm}
\psfig{figure=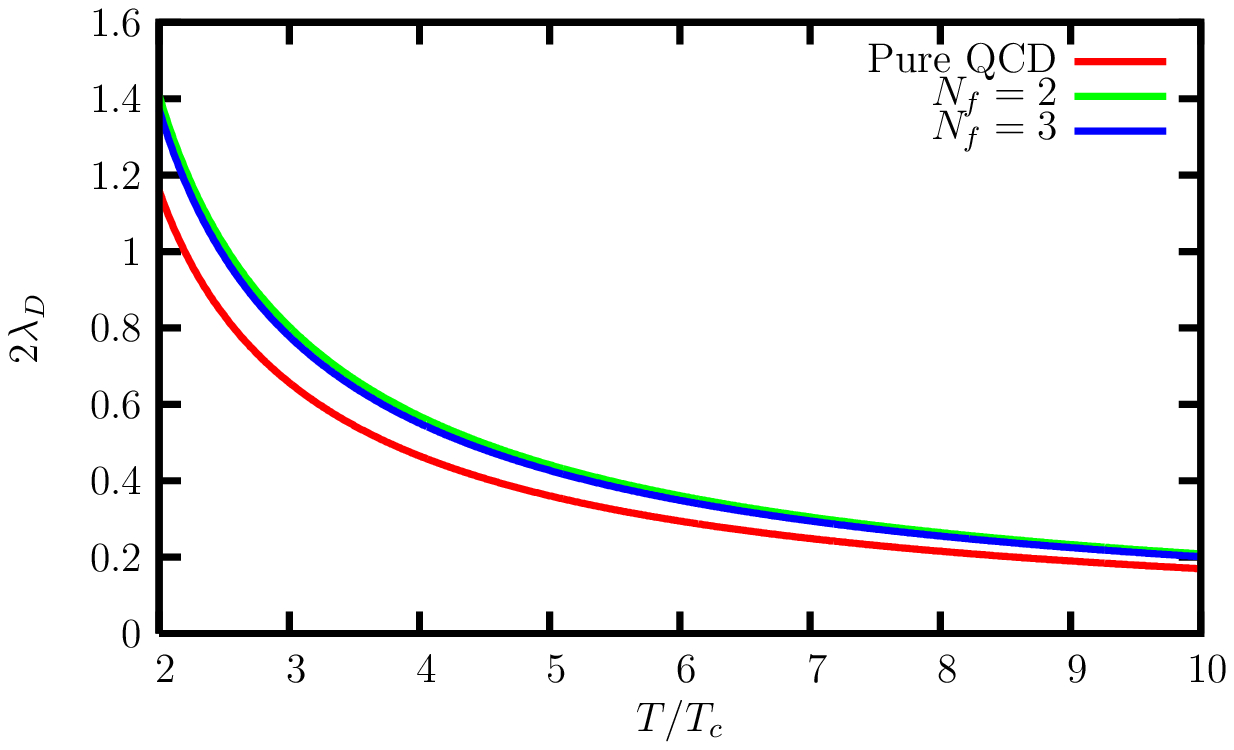,width=100mm}
\hspace{-30mm}
\psfig{figure=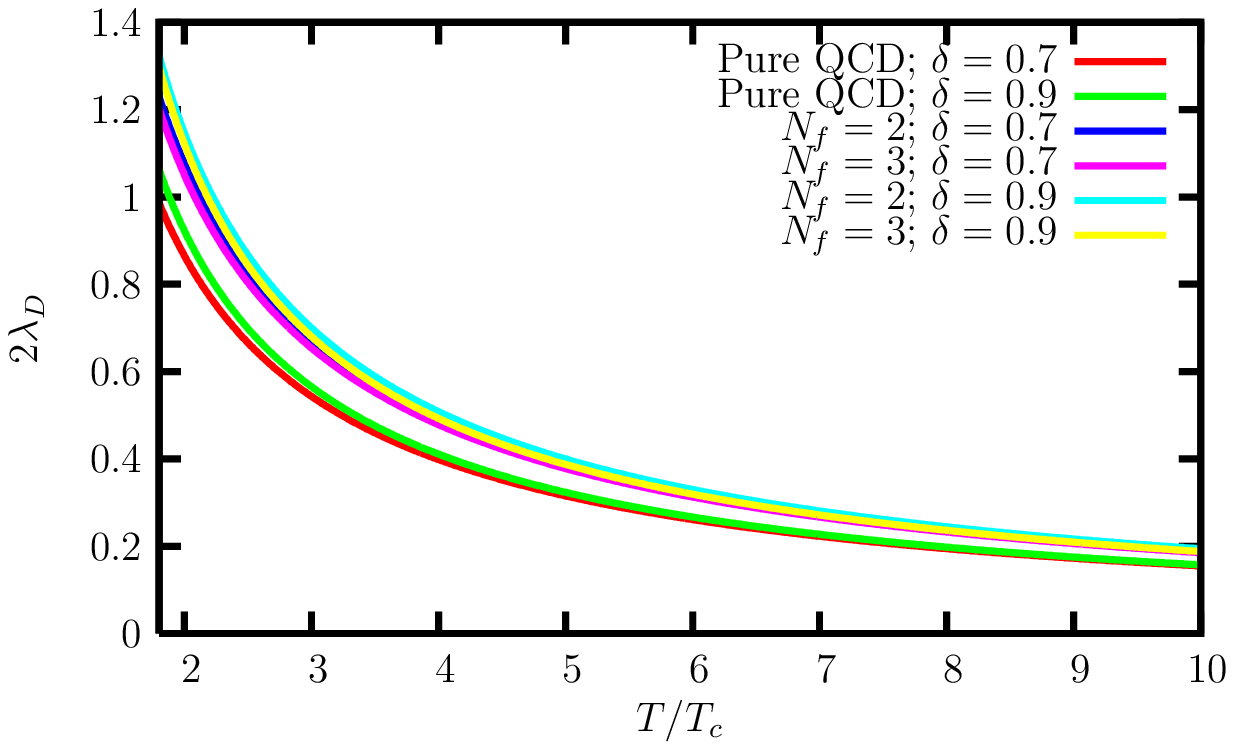,width=100mm}
\vspace*{-45mm}
\caption{Twice of the Debye screening length $\lambda_D$ is plotted  as function of $T/T_c$ for Pure QCD, $N_f=2$and $N_f=3$ for EOS1 and EOS2. Please note that the phrase Pure QCD is used for pure gauge theory. The left Fig. is for EOS1 and right one is for EOS2. Note that the screening length has been measured in {\it $fm^{-1}$}.}
\end{figure}

\section{The heavy quark potential and dissociation of quarkonia}
In this section, we address the screening of heavy quark potential by 
%It is well known from QED that the medium modifications to interaction potential between the two unlike static charges
%enters in to the Fourier transform of the potential through the dielectric permittivity. The same philosophy applies for QGP.
 combining the quasi-particle model for  hot QCD introduced in Section(I) to the semi-classical transport theory\cite{aranjan}. 
The response function (chromo-electric permittivity) has recently been determined 
from the transport theory by employing the ideal distribution functions for gluons and quarks by 
Ravishankar and Ranjan\cite{aranjan}. We determine these response functions for realistic distribution functions obtained from 
pQCD. Finally, the response functions is then used to study the screening properties of QGP. We outline the method(discussed in detail in\cite{aranjan,chandra1,chandra2}) below.

 Consider the Cornell potential: $\phi(r)=-\frac{\alpha}{r}+\Lambda r$, where $\alpha$ and $\Lambda$ are  phenomenological constants. 
The first term  dominates at small distance while the linear causes confinement,  dominating at large distances. 
The medium modifies the expression for the potential (in the Fourier space),
$\tilde{\phi}(k) \rightarrow \tilde{\phi}(k)/\tilde{\epsilon}(k)$, where the response is evaluated at
$\omega=0$. Accordingly, the  potential undergoes  modification $\tilde{\phi}(k) \rightarrow \tilde{\phi}_m(k)$, 
which can be written as 
\begin{eqnarray}
\label{eqn17}
\tilde{\phi}_m(k)=-\sqrt{\frac{2}{\pi}}\frac{\alpha}{k^2+m^2_D}
-\frac{4}{\sqrt{2\pi}}\frac{\Lambda }{k^2(k^2+m^2_D)},
\end{eqnarray}
where
\begin{eqnarray}
\label{eqn19}
m^2_D=8\pi Q^2T^2\bigg[N_f f_2(z_q) +g_2(z_g)\bigg],
\end{eqnarray}
For a gluonic plasma, the above expression reduces to
$m^2_{Dg}=16\pi  Q^2T^2 g_{2}(z_g)$, where $z_g=\exp(\tilde\mu_g)$ and $z_q=\exp(\tilde\mu_q)$ Note that, $m_D$ has determined by  the permittivity via:
$\epsilon=1+ \frac{m^2_D}{k^2}$. The modified form of the potential in r-space is then given by:
\begin{eqnarray}
\label{eqn21}
\phi_s(r)&=&(\frac{2\Lambda}{m^2_D}-\alpha)\frac{\exp{(-m_Dr)}}{r}
-\frac{2\Lambda}{m^2_Dr}+\frac{2\Lambda}{m_D}-\alpha m_D.
\end{eqnarray}

It is clear from Eq.(6) that the  role played by the Debye mass in QGP is rather different from its role in electrodynamic plasma.
To see this, note that at large $T$, Eq.(6) reduces to
\begin{eqnarray}
\label{lrp}
\phi_s(r)\sim -\frac{2\Lambda}{m^2_Dr}-\alpha m_D
\end{eqnarray}

Ignoring the additive contribution, the  energy of the $q\bar{q}$ in the ground state is simply given
by
\begin{equation} 
E_g=\frac{m_q\Lambda^2}{m^4_D},
\end{equation}

where $m_q$ is the mass of heavy quark. The binding energy is, of course, temperature dependent and approaches zero as $T \rightarrow \infty$. At any finite temperature though, 
the quarks possess a thermal energy $E_{Th} \sim \frac{3}{2} T$, by equipartition theorem,
 leading to an ionization of the quarkonium when $E_{Th}$ matches the binding energy. This leads to the  melting temperature $T_d$ of $J/\Psi$ and $\Upsilon$ listed in Table(I).
\begin{table}
\label{table1}
\caption{The dissociation temperature($T_D$) for various quarkonia states (in unit of $T_c$).}
\centering
\begin{tabular}{|l|l|l|l|l|}
\hline
Hot EOS& quarkonium &Pure QCD& $N_f=2$&$N_f=3$\\
&&&&\\
\hline\hline
EOS1&$J/\Psi$&2.2&2.62&2.46\\
&$\Upsilon$&2.5&3.14&2.94\\
\hline
EOS2&$J/\Psi$&1.86&2.38&2.24\\
$\delta=0.8$& $\Upsilon$&2.12&2.76&2.58\\
\hline
EOS2&$J/\Psi$&1.95&2.45&2.32\\
$\delta=1.0$&$\Upsilon$&2.2&2.83&2.66\\
\hline
EOS2&$J/\Psi$&2.03&2.52&2.40\\
$\delta=1.2$&$\Upsilon$&2.28&2.9&2.74\\
\hline
\end{tabular}
\end{table}
It is noteworthy that the dissociation temperatures are all roughly in the range $ T_D \approx (2-3)T_c$, which is higher than the temperatures achieved so far. Since the temperatures expected at LHC is $\sim 2T_c-3T_c$, one may expect to test these predictions there. And moreover these estimates are consistent with the recent results from other theoretical works\cite{disso}.
\section{Summary and conclusions}
In conclusion, we have developed a quasi particle model for hot QCD to extract the equilibrium distribution functions 
for gluons and quarks from the hot QCD EOS. We have shown that the interaction effects can entirely be captured 
in the effective fugacities for the gluons and quarks. We utilized these distribution functions to study 
the screening length as a function of temperature. We have determined the dissociation temperatures for $J/\Psi$ and $\Upsilon$ by studying the medium modifications
to heavy quark potential which enters in the form of chromo-electric permittivity. The results on dissociation temperatures are consistent 
with other recent theoretical works. Our approach can easily be generalized to the hot QCD EOS at finite quark-chemical potential. This approach can also be employed to study the thermodynamic and transport properties of hot and
dense matter at RHIC.

\vspace{2mm}
\noindent {\bf Acknowledgment}: VC acknowledges Akhilesh Ranjan for fruitful discussions. We are grateful to the people of India for their valuable support for the research in basic sciences. VC also acknowledges CSIR New Delhi, India for the financial support.


\begin{thebibliography}{50}
\bibitem{star} STAR collaboration, J. Adams {\it et al.}
Nucl. Phys. A {\bf 757} 102(2005).
\bibitem{weakc} Anton Rebhan,{\tt hep-ph/0504023} and references therein.
\bibitem{chandra1} Vinod Chandra, Ravindra Kumar and V. Ravishankar,
Phys. \ Rev. {\bf C 76} (2007) 054909.
\bibitem{chandra2} Vinod Chandra, Akhilesh Ranjan and V. Ravishankar,
{\tt arXiv:0801.1286(hep-ph)}.
\bibitem{fkarsch} Frithjof Karsch,
Lect.\ Notes \ Phys. {\bf 583} (2002) 209 ({\tt arXiv:hep-lat/0106019})(Please see Fig.13 of this Ref.) 
\bibitem{aranjan} Akhilesh Ranjan and V. Ravishankar, {\tt arXiv:0708.3697(nucl-th)}.
\bibitem{arnold} P. Arnold and Chengxing Zhai, Phys. Rev. D {\bf 50}  7603(1994); Phys. Rev. D {\bf 51}  1906(1995);
Chengxing Zhai and B. Kastening, Phys. Rev. D {\bf 52}  7232(1995).
\bibitem{kaj1} K. Kajantie, M. Laine, K. Rummukainen and Y. Schroder,
Phys. Rev. D {\bf 67} 105008(2003).
\bibitem{kelly}  P. F. Kelly, Q. Liu, C. Lucchesi and C. Manual,
Phys.\ Review. \ Lett. {\bf 72}, 3461 (1994); Phys.\ Rev. {\bf D 50}, 4209(1994).
\bibitem{zantow} O. Kazmarek and F. Zantow, 
POS(LAT-2005) 177.
\bibitem{disso} Helmut Satz, Nucl.\ Phys. {\bf A 783}, (2007) 249;  W. M. Alberico, A. Beraudo, A. De Pace and A. Molinari,
Phys.\ Rev. {\bf D 75}, 074009 (2007).

\end{thebibliography}
\end{document}